# Perfect Diffraction with Bianisotropic Metagratings

Zhiyuan Fan[1], Maxim R. Shcherbakov[1], Monica Allen[2], Jeffery Allen[2], and Gennady Shvets[1]

[1] School of Applied and Engineering Physics, Cornell University, Ithaca NY 14853, USA

[2] Air Force Research Laboratory, Munitions Directorate, Eglin FL 32542, USA

**Abstract**

One highly desirable function of a diffraction grating is its ability to deflect incident light into a specific diffraction order with near-perfect efficiency. While such asymmetry can be achieved in a variety of ways, e.g., by using a sawtooth (blazed) geometry, a recently emerged approach is to use a planar metagrating comprised of designer multi-resonant periodic units (metamolecules). Here we demonstrate that a bianisotropic unit cell supporting four resonances interfering in the far field can be used as a building block for achieving the prefect deflection. A coupled mode analysis shows that these modes provide a small number of orthogonal electromagnetic radiation patterns that are needed to suppress transmission/reflection into all but one diffraction order. Bianisotropy caused by a mirror symmetry breaking enables a normally incident wave to excite, through near-field couplings, two otherwise "dark" resonant modes. We design and experimentally realize bianisotropic metamolecules which are sub-wavelength in all three dimensions, and whose optical properties are desensitized to fabrication imperfections by their geometric simplicity. We show that optical beams tightly focused onto the metagratings with just a few unit cells can also be asymmetrically deflected with high efficiency, paving the way for compact broadband optical devices.

# Introduction

High-refractive-index all-dielectric metastructures [ 1, 2, 3, 4, 5] are well-known for their negligible non-radiative losses, geometry-dependent Mie resonances [ 6, 7], and high field enhancements that translate into exciting opportunities in nonlinear optics [ 8, 9, 10]. Their ability to support multiple resonances has been shown to be crucial for asymmetric light scattering. For example, the overlap between in-plane electric and magnetic dipole [ 11, 12, 13, 14, 15, 16] resonances has been shown to produce an asymmetry between the forward and backward scattering. The most extreme manifestation of the interference between the resonances is a reflection-free (Huygens) metasurface [ 17]. Such metasurfaces have been used for controlling the phase of the transmitted radiation [ 18, 19] and other optical applications such as ultra-thin metalenses, holograms, and polarization control [ 20, 21, 22, 23, 24].

More recently, ever more complex resonances of metamolecules comprising a metasurface have been explored. Of particular interest are asymmetrically coupled resonances that impart metamolecules with finite bianisotropy [ 25, 26, 27]: an electromagnetic property that enables the excitation of magnetic resonances using an electric field. Individual and arrayed bianisotropic metamolecules have been designed and fabricated using a variety of material platforms, both dielectric and metallic, across a wide spectral range, from centimeters to nanometers [ 4, 28, 29, 30, 31, 32, 33, 34, 35]. While the interference between *in-plane* electric and magnetic resonances excited by a normally incident electromagnetic wave can introduce the up-down (forward/backward) scattering asymmetry, it has recently been shown that bianisotropic coupling between *in-plane* electric and *out-of-plane* magnetic moments can induce strong lateral scattering asymmetry [ 28, 36] by a single metamolecule.

Despite the wide interest in multi-resonant metamolecules[ 37] with in- and out- of the plane electric/magnetic resonances, there has not been a systematic study of the possible functionalities of such metamolecules depending on the number and the nature of the resonant modes. Because scattering by a single sub-wavelength metamolecule is small, the most dramatic effects are expected for metamolecules arranged as a metasurface. Moreover, because of their resonant nature, dielectric metasurfaces are particularly appealing because their non-radiative losses can be negligibly small. Below, we demonstrate that an array of bianisotropic metamolecules comprising a bianisotropic metasurface can be employed as a "perfect" diffraction grating: an optical element that deflects a normally incident electromagnetic wave into a single diffractive order, without losses to reflection or to other diffractive orders. The array is assumed to be sub-wavelength in one lateral direction (i.e. its period $P_y$ satisfies $P_y < \lambda$, where $\lambda$ is the wavelength of light), and super-wavelength in the other direction ($P_x > \lambda$), with the diffractive orders emerging in the $x - z$ plane, two in transmission and two in reflection. We theoretically demonstrate that such a metasurface must possess at least 4 coupled resonances to be perfectly efficient, when an additional scattering asymmetry is enabled other than the up-down- and left-right-asymmetries that have been extensively studied previously. A bianisotropic design of a Si-based metasurface is experimentally investigated, and the high diffraction efficiency is confirmed for mid-infrared (mid-IR) light. Finally, we demonstrate that such a low-order metasurface can provide efficient diffraction even if fabricated in small patches. Such an ultrathin metasurface presents a new design paradigm for ultimately efficient gratings capable of large-angle deflection over a fairly broad bandwidth.

Many design approaches to making high-efficiency gratings have been tried over the years, including sawtooth-shaped gratings or, more recently, flat metasurfaces [ 20, 22, 23, 38, 39, 40] consisting of waveguiding pillars of constant thickness $h$. These designs can produce nearly arbitrary gradients of phase retardation, and can be used for a variety of applications, including holograms, wavefront shaping, and meta-lensing applications. While the simplicity of the pillar-based designs is very appealing, their performance is limited by several factors, including Fresnel reflections, the need to use fairly thick structures ($h \sim h_s = \lambda/(n-1)$, where $n$ is the refractive index of the pillars), and a relatively small number of free geometric parameters of a unit cell of a metasurface that can be used for performance optimization. As the result, the absolute diffraction efficiency, defined as the fraction of the incident energy flux transmitted into the desired diffractive order, is typically about 80% at best [ 39, 41, 42]. More recently, a new design strategy based on the inverse design approach has been shown [ 43] to provide diffraction efficiencies that in theory approach 90%. The main challenge of using this approach is the non-intuitive nature of the design that could potentially create over exploration of geometric dimensions, difficulty in accurate fabrication of fine features as well as diffraction angle scaling. Therefore, there is a need for a design strategy that uses simple geometric shapes of the metamolecules, yet retains a sufficient number of free geometric parameters to ensure high efficiency. Below, we present a minimal theoretical model of an efficient diffractive grating that functions as a beam deflector, and is based on a bianisotropic metasurface that supports simple and intuitive electromagnetic resonances.

## Results and Discussion

**Coupled-mode theory.** Electromagnetic (EM) resonances of metamolecules that were selected to fulfill the blazed grating condition are illustrated in Fig.1(a). In general, a modal electric field profile $\mathbf{E}$ can be decomposed into far-field components $\mathbf{E}_{\mathrm{rad}}$ and near-field components $\mathbf{E}_{\mathrm{loc}}$. The near fields are confined to the metasurface. Here, we assume the dielectric medium is lossless and the localized near fields generate zero Ohmic heat. The far fields are radiative components which take away EM energy in the form of plane wave diffractions. For a specific modal profile, $\mathbf{E}(\mathbf{k}_\mathbf{B}, \mathbf{r}) = \{E_x(\mathbf{k}_\mathbf{B}, \mathbf{r}), E_y(\mathbf{k}_\mathbf{B}, \mathbf{r}), E_z(\mathbf{k}_\mathbf{B}, \mathbf{r})\}$, where $\mathbf{k}_\mathbf{B}$ defines the Bloch vector for the mode and $\mathbf{r}$ defines the coordinate 3-vector $\{x, y, z\}$ within the unit cell of the metasurface, the amplitude of a diffraction beam with indices $(m, n)$ can be found as an overlap integral between the modal profile and a Fourier component on the top or bottom interface of a metasurface, following a plane wave expansion of the transverse component of the electric field [44, 45]. In the case of an $s$-polarized wave,

$$\alpha_j^{(m,n),\pm} \equiv \int_0^{P_y} \int_0^{P_x} \mathbf{E}_j\left(\mathbf{k}_\mathbf{B}, \left\{x, y, \pm\frac{h}{2}\right\}\right) \cdot \mathbf{e}_s^{(m,n)} e^{-i(k_x x + k_y y)} e^{-i(mG_x x + nG_y y)} dx dy, \quad (1)$$

where the integral is performed on the metasurface's interfaces located at $\pm h/2$ within a unit cell. A unit vector $\mathbf{e}_s^{(m,n)}$ defines the transverse electric field component of an $s$-polarized diffractive order corresponding to the indices $(m, n)$. Note that the Bloch vector is defined by a the $x$- and $y$-components of the incident wavevector $\mathbf{k}_\mathrm{B} = \{k_x, k_y\}$, and $\frac{\omega}{c} = \sqrt{k_x^2 + k_y^2 + k_z^2}$ is the amplitude of the incident wave vector in vacuum. A mode profile usually is also a function of discrete resonance frequencies $\omega_j(\mathbf{k}_\mathrm{B})$, where $j$ denotes the mode index here and also in Eq.(1). $G_x = 2\pi/P_x$ and $G_y = 2\pi/P_y$ are wavevectors of the reciprocal lattice of the metamolecule array or the metagrating. For a diffractive beam with order indices $(m, n)$, it requires that $\frac{\omega^2}{c^2} > (k_x + mG_x)^2 + (k_y + nG_y)^2$. Otherwise, the overlap integral finds a coefficient for a vertically evanescent near field component. In the discussion, we have assumed that the ambient medium is air with a refractive index of one. Also, **E** can be normalized according to a total near field energy, i.e. $\iint \varepsilon |\mathbf{E}_{\text{loc}}|^2 dV = W_0$. For diffractive orders that are p-polarized, Eq.(1) can be conveniently adapted. One way is to calculation overlap integrals using magnetic fields that are parallel to the interfaces.

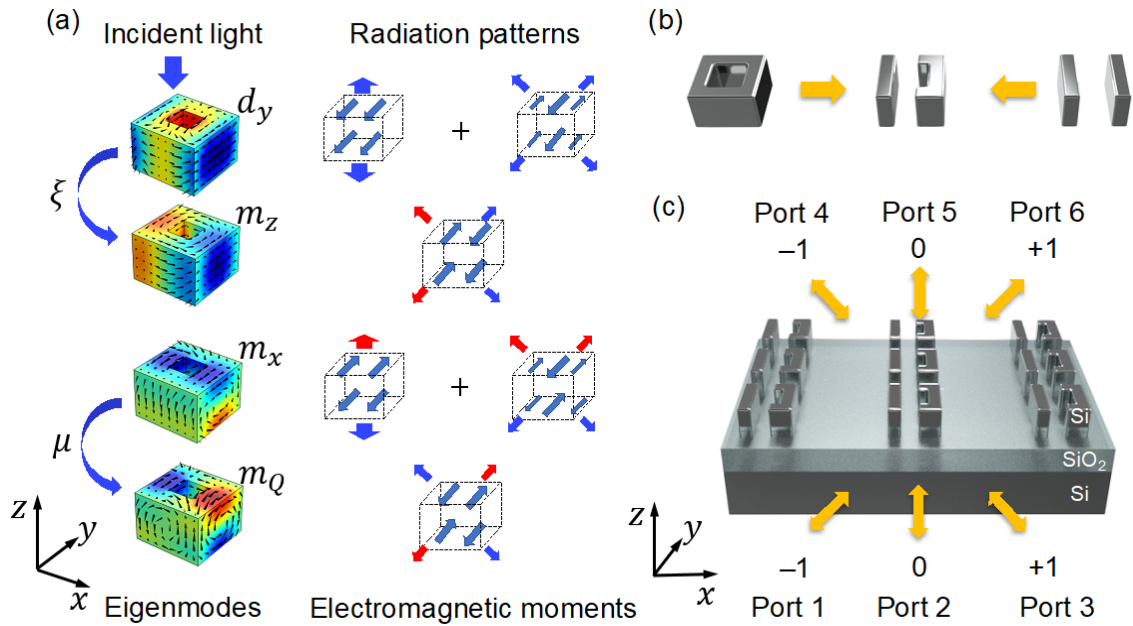

**Figure 1.** (a) Illustration of eigenmodes and radiation of corresponding electromagnetic moments. Four fundamental electromagnetic modes are illustrated by the colored rectangular rings, in which amplitude of $E_y$ is color-coded, and the cones represent the direction of the full field. Each mode has signature EM field components as shown next to these eigenmodes. The boxes depict volumes of abstract meta-atoms within a unit cell. The arrows inside boxes indicate displacement current distributions. The arrows outside the boxes illustrate diffraction patterns when these resonators are assembled into a grating. The phase of the diffracting fields is color-coded, with red and blue being out of phase. An electric mode ($d_y$) and a magnetic mode ($m_x$) usually have two components each: one component radiates normally with respect to the surface and the other component radiates symmetrically into ports on the left and right. Coupling coefficients $\xi$ and $\mu$ between corresponding modes are non-zero when bianisotropy is introduced. (b) Bianisotropy can be introduced by breaking the mirror symmetry about the y-axis as shown by the silicon '| |'-shaped

metamolecule in the middle. (c) A schematic of a 6-port metasurface. This figure depicts our device designed and fabricated out of a silicon on insulator wafer.

With the definition of Eq.(1), we can conveniently find out not only the amplitudes, but also the phases of diffraction beams radiated by a mode. Considering a group of EM resonances of various symmetries, vectors consisting of complex amplitudes of diffractive waves at each port may define a basis for diffraction patterns. Within the scope of this manuscript, we further assume the incident electric field is $y$-polarized, propagates in the normal direction to the metasurface, i.e., $k_x = 0, k_y = 0$, and the metagrating only diffracts along the $x$-direction with $+1^{st}$ and $-1^{st}$ orders, which defines a 6-port diffracting system for $y$-polarization as shown in Fig.1(c). This way, the indices of a diffraction beam in Eq.(1) are simplified. For an electric mode $d_y$, its electric field profile is symmetric about both $zy$–plane and $xy$–plane. Therefore, $\alpha_{d_y}^{(2)} = \alpha_{d_y}^{(5)} \equiv \alpha_{d_y}$ and $\alpha_{d_y}^{(1)} = \alpha_{d_y}^{(3)} = \alpha_{d_y}^{(4)} = \alpha_{d_y}^{(6)} \equiv \alpha'_{d_y}$. For an *in-plane* magnetic mode $m_x$, its electric field profile is symmetric about $zy$–plane and anti-symmetric about the $xy$–plane. Therefore, $\alpha_{m_x}^{(5)} = -\alpha_{m_x}^{(2)} \equiv \alpha_{m_x}$ and $\alpha_{m_x}^{(4)} = \alpha_{m_x}^{(6)} = -\alpha_{m_x}^{(1)} = -\alpha_{m_x}^{(3)} \equiv \alpha'_{m_x}$. For an *out-of-plane* magnetic mode $m_z$, its electric field profile is anti-symmetric about $zy$– but symmetric about the $xy$–plane. Therefore, $\alpha_{m_z}^{(2)} = \alpha_{m_z}^{(5)} = 0$ and $\alpha_{m_z}^{(3)} = -\alpha_{m_z}^{(1)} = \alpha_{m_z}^{(6)} = -\alpha_{m_z}^{(4)} \equiv \alpha'_{m_z}$. The last mode that we will discuss is referred to as the magnetic quadrupole mode $m_Q$. This mode is anti-symmetric about both $zy$– plane and $xy$–plane. Then, $\alpha_{m_Q}^{(2)} = \alpha_{m_Q}^{(5)} = 0$ and $\alpha_{m_Q}^{(1)} = -\alpha_{m_Q}^{(3)} = -\alpha_{m_Q}^{(4)} = \alpha_{m_Q}^{(6)} \equiv \alpha'_{m_Q}$. This prescription can be conveniently generalized for a metagrating with any number of diffraction ports. Note that for EM modes excited under oblique incidence, one needs to use the Bloch solutions in the form of $\mathbf{E}(\mathbf{r})e^{-i(k_x x + k_y y)}$.

In a metagrating, energy transfer among EM modes and diffraction ports can be phenomenologically investigated using a coupled-mode theory (CMT) model [46, 47, 48]. A multi-port CMT theory is now adapted for the 6-port metagrating as defined previously and for the $y$-polarized light :

$$\frac{d}{dt}\vec{A} = \begin{pmatrix} -i\widetilde{\omega}_{d_y} & i\xi & & 0 \\ i\xi & -i\widetilde{\omega}_{m_z} & & \\ & & -i\widetilde{\omega}_{m_x} & i\mu \\ 0 & & i\mu & -i\widetilde{\omega}_{m_Q} \end{pmatrix} \cdot \vec{A} + \bar{\bar{\beta}} \cdot \vec{S}_+, \tag{2}$$

where $\vec{A} \equiv \{d_y, m_z, m_x, m_Q\}^T$ defines the amplitudes of the EM modes; $\vec{S}_+ \equiv \{S_+^1, S_+^2, S_+^3, S_+^4, S_+^5, S_+^6\}^T$ defines the amplitudes of in-coming waves; $\vec{S}_+ = \{0, e^{-i\omega t}, 0, 0, 0, 0\}^T$ is assumed for the current problem, and $\omega$ is the excitation frequency; $\widetilde{\omega}_j$ is the complex frequency of a resonating EM mode $j$, with its real part indicating a resonance frequency and imaginary part indicating an energy loss rate; $\xi$ and $\mu$ are two non-vanishing inter-mode coupling coefficients resulting from the symmetry breaking of a metamolecule illustrated in Fig.1(a) and (b); matrix elements $\bar{\bar{\beta}}_j^{\,i}$ are the coupling coefficients for an external field at port $i$ to excite a mode $j$. After mode amplitudes are known, the amplitudes of diffractive waves are found using the following equation:

$$\vec{S}_- = \bar{\bar{C}} \cdot \vec{S}_+ + \bar{\bar{\alpha}} \cdot \vec{A}. \tag{3}$$

Here, $\vec{S}_- \equiv \{S_-^1, S_-^2, S_-^3, S_-^4, S_-^5, S_-^6\}^T$, and $S_-^j$ are the amplitudes of out-going waves at each port. Matrix elements $\bar{\bar{\alpha}}^i{}_j$ are the coupling coefficients from mode $j$ to diffraction port $i$. $\bar{\bar{C}}$ is the so-called direct coupling coefficient matrix [47] and it takes the form as shown in the following for this study.

$$\bar{\bar{C}} = \begin{pmatrix} 0 & 0 & 0 & 0 & 0 & 1 \\ 0 & 0 & 0 & 0 & 1 & 0 \\ 0 & 0 & 0 & 1 & 0 & 0 \\ 0 & 0 & 1 & 0 & 0 & 0 \\ 0 & 1 & 0 & 0 & 0 & 0 \\ 1 & 0 & 0 & 0 & 0 & 0 \end{pmatrix}. \tag{4}$$

We can prescribe all elements of $\bar{\bar{\alpha}}$ using our previous derivation of diffraction field amplitudes for each mode. Note that these coefficients $\bar{\bar{\alpha}}^i{}_j$ will differ from Eq.(1) only by a diffraction-angle-dependent normalization factor, when $|S_-^i|^2$ are normalized as diffraction efficiencies under a unity incident field $|S_+^2| = 1$. It is required by time reversal symmetry and energy conservation [47, 48] that

$$\mathrm{diag}(\bar{\bar{\alpha}} \cdot \bar{\bar{\alpha}}^T) = 2\left(\mathrm{Im}\tilde{\omega}_{d_y}, \mathrm{Im}\tilde{\omega}_{m_z}, \mathrm{Im}\tilde{\omega}_{m_x}, \mathrm{Im}\tilde{\omega}_{m_Q}\right) \tag{5}$$

and

$$\bar{\bar{\alpha}} = \bar{\bar{\beta}}^T = \begin{pmatrix} \tilde{\alpha}'_{d_y} & \tilde{\alpha}'_{m_z} & -\tilde{\alpha}'_{m_x} & \tilde{\alpha}'_{m_Q} \\ \tilde{\alpha}_{d_y} & 0 & -\tilde{\alpha}_{m_x} & 0 \\ \tilde{\alpha}'_{d_y} & -\tilde{\alpha}'_{m_z} & -\tilde{\alpha}'_{m_x} & -\tilde{\alpha}'_{m_Q} \\ \tilde{\alpha}'_{d_y} & \tilde{\alpha}'_{m_z} & \tilde{\alpha}'_{m_x} & -\tilde{\alpha}'_{m_Q} \\ \tilde{\alpha}_{d_y} & 0 & \tilde{\alpha}_{m_x} & 0 \\ \tilde{\alpha}'_{d_y} & -\tilde{\alpha}'_{m_z} & \tilde{\alpha}'_{m_x} & \tilde{\alpha}'_{m_Q} \end{pmatrix}. \tag{6}$$

The tilda introduced in Eq.(6) is nothing but a reminder of the normalization factor. Then, we may solve Eqs.(2) and (3) with the coupling coefficients defined in Eq.(6). We obtained the diffraction coefficients for each port under a *y*-polarized normal incident light.

$$t_1 = i\left\{-\frac{\tilde{\alpha}_{d_y}\left(\tilde{\alpha}'_{d_y}\Delta\tilde{\omega}_{m_z} - \tilde{\alpha}'_{m_z}\xi\right)}{\Delta\tilde{\omega}_{d_y}\Delta\tilde{\omega}_{m_z} - \xi^2} + \frac{\tilde{\alpha}_{m_x}\left(\tilde{\alpha}'_{m_x}\Delta\tilde{\omega}_{m_Q} + \tilde{\alpha}'_{m_Q}\mu\right)}{\Delta\tilde{\omega}_{m_x}\Delta\tilde{\omega}_{m_Q} - \mu^2}\right\},$$

$$t_2 = -i\left\{\frac{\left(\tilde{\alpha}_{d_y}\right)^2 \Delta\tilde{\omega}_{m_z}}{\Delta\tilde{\omega}_{d_y}\Delta\tilde{\omega}_{m_z} - \xi^2} - \frac{\left(\tilde{\alpha}_{m_x}\right)^2 \Delta\tilde{\omega}_{m_Q}}{\Delta\tilde{\omega}_{m_x}\Delta\tilde{\omega}_{m_Q} - \mu^2}\right\},$$

$$t_3 = i\left\{-\frac{\tilde{\alpha}_{d_y}\left(\tilde{\alpha}'_{d_y}\Delta\tilde{\omega}_{m_z} + \tilde{\alpha}'_{m_z}\xi\right)}{\Delta\tilde{\omega}_{d_y}\Delta\tilde{\omega}_{m_z} - \xi^2} + \frac{\tilde{\alpha}_{m_x}\left(\tilde{\alpha}'_{m_x}\Delta\tilde{\omega}_{m_Q} - \tilde{\alpha}'_{m_Q}\mu\right)}{\Delta\tilde{\omega}_{m_x}\Delta\tilde{\omega}_{m_Q} - \mu^2}\right\},$$

$$t_4 = -i\left\{\frac{\tilde{\alpha}_{d_y}\left(\tilde{\alpha}'_{d_y}\Delta\tilde{\omega}_{m_z} - \tilde{\alpha}'_{m_z}\xi\right)}{\Delta\tilde{\omega}_{d_y}\Delta\tilde{\omega}_{m_z} - \xi^2} + \frac{\tilde{\alpha}_{m_x}\left(\tilde{\alpha}'_{m_x}\Delta\tilde{\omega}_{m_Q} + \tilde{\alpha}'_{m_Q}\mu\right)}{\Delta\tilde{\omega}_{m_x}\Delta\tilde{\omega}_{m_Q} - \mu^2}\right\},$$

$$t_5 = 1 - i\left\{\frac{\left(\tilde{\alpha}_{d_y}\right)^2 \Delta\tilde{\omega}_{m_z}}{\Delta\tilde{\omega}_{d_y}\Delta\tilde{\omega}_{m_z} - \xi^2} + \frac{\left(\tilde{\alpha}_{m_x}\right)^2 \Delta\tilde{\omega}_{m_Q}}{\Delta\tilde{\omega}_{m_x}\Delta\tilde{\omega}_{m_Q} - \mu^2}\right\},$$

$$t_6 = -i\left\{\frac{\tilde{\alpha}_{d_y}\left(\tilde{\alpha}'_{d_y}\Delta\tilde{\omega}_{m_z} + \tilde{\alpha}'_{m_z}\xi\right)}{\Delta\tilde{\omega}_{d_y}\Delta\tilde{\omega}_{m_z} - \xi^2} + \frac{\tilde{\alpha}_{m_x}\left(\tilde{\alpha}'_{m_x}\Delta\tilde{\omega}_{m_Q} - \tilde{\alpha}'_{m_Q}\mu\right)}{\Delta\tilde{\omega}_{m_x}\Delta\tilde{\omega}_{m_Q} - \mu^2}\right\}.$$

$$\tag{7}$$

By taking the limit of $\Delta\tilde{\omega}_j \equiv \tilde{\omega}_j - \omega \to \infty$, we can exclude the $j$-th mode from the CMT model, which allows us to investigate the best efficiency of diffraction into the 6-th port when not all of the 4 EM modes are active. Analytical solutions can be found for cases when only $d_y$ and/or $m_x$ modes are active. Utilizing $\text{Im}\tilde{\omega}_{d_y/m_x} = (\tilde{\alpha}_{d_y/m_x})^2 + 2(\tilde{\alpha}'_{d_y/m_x})^2$ from Eq.(4), we may find $|S^6_-|_{max} = 12.5\%$ when only one of the modes is excited, and 50% when both of the modes are excited. For the case with $d_y$ and $m_z$, the best diffraction efficiency is found numerically to be 25%, whereas the maximum efficiency is 73% for the case with $d_y$, $m_z$ and $m_x$. A near-perfect, 99.8% diffraction efficiency was found when all of the four modes are active; see Supporting Information part 7 for details. These results establish that 4 modes of various symmetries is the minimum requirements needed for a perfect 6-port diffraction grating, whereas using less fundamental modes will result in inefficient diffraction.

We can also approach this problem by enforcing all $|S^i_-|$ but $|S^6_-|$ to be 0. This leads to one possibility presented by Eqs.(8-10).

$$\tilde{\alpha}_{d_y}\tilde{\alpha}'_{d_y} - \frac{\tilde{\alpha}_{d_y}\tilde{\alpha}'_{m_z}\xi}{\Delta\tilde{\omega}_{m_z}} \approx 0,$$

$$\tilde{\alpha}_{m_x}\tilde{\alpha}'_{m_x} + \frac{\tilde{\alpha}_{m_x}\tilde{\alpha}'_{m_Q}\mu}{\Delta\tilde{\omega}_{m_Q}} \approx 0,$$

(8)

and

$$\frac{\tilde{\alpha}_{d_y}(\tilde{\alpha}'_{d_y}\Delta\tilde{\omega}_{m_z} + \tilde{\alpha}'_{m_z}\xi)}{\Delta\tilde{\omega}_{d_y}\Delta\tilde{\omega}_{m_z} - \xi^2} - \frac{\tilde{\alpha}_{m_x}(\tilde{\alpha}'_{m_x}\Delta\tilde{\omega}_{m_Q} - \tilde{\alpha}'_{m_Q}\mu)}{\Delta\tilde{\omega}_{m_x}\Delta\tilde{\omega}_{m_Q} - \mu^2}$$

$$= \frac{2\tilde{\alpha}_{d_y}\tilde{\alpha}'_{m_z}\xi}{\Delta\tilde{\omega}_{d_y}\Delta\tilde{\omega}_{m_z} - \xi^2} + \frac{2\tilde{\alpha}_{m_x}\tilde{\alpha}'_{m_Q}\mu}{\Delta\tilde{\omega}_{m_x}\Delta\tilde{\omega}_{m_Q} - \mu^2} \approx 0.$$

(9)

The physical meaning of Eqs.(8) is that fields radiated by the dark modes ($m_z$ and $m_Q$) cancel the fields from the bright modes ($d_y$ and $m_x$, respectively) in port 1 and 4. While Eq.(9) implies that fields radiated by $d_y$ and $m_z$ are in phase in port 3, but their sum is out of phase by $\pi$ with respect to the total field of $m_x$ and $m_Q$. At this point, we observe a phase relation among the radiative patterns of these fundamental modes. This is one of the necessary conditions for perfect diffraction efficiency into the 6th port. Another condition requires both 0th order reflection and transmission (ports 2 and 5) to vanish, which is found from the following relations:

$$S_-^2 = -i \left\{ \frac{(\tilde{\alpha}_{m_x})^2}{\Delta\tilde{\omega}_{m_x} - \frac{\mu^2}{\Delta\tilde{\omega}_{m_Q}}} - \frac{(\tilde{\alpha}_{d_y})^2}{\Delta\tilde{\omega}_{d_y} - \frac{\xi^2}{\Delta\tilde{\omega}_{m_z}}} \right\} \approx 0,$$

$$S_-^5 = 1 - i \left\{ \frac{(\tilde{\alpha}_{m_x})^2}{\Delta\tilde{\omega}_{m_x} - \frac{\mu^2}{\Delta\tilde{\omega}_{m_Q}}} + \frac{(\tilde{\alpha}_{d_y})^2}{\Delta\tilde{\omega}_{d_y} - \frac{\xi^2}{\Delta\tilde{\omega}_{m_z}}} \right\} \approx 0.$$

(10)

They can be simplified down to $\frac{i(\tilde{\alpha}_{m_x})^2}{\Delta\tilde{\omega}_{m_x} - \mu^2/\Delta\tilde{\omega}_{m_Q}} = \frac{i(\tilde{\alpha}_{d_y})^2}{\Delta\tilde{\omega}_{d_y} - \xi^2/\Delta\tilde{\omega}_{m_z}} \approx 1/2$. Previously in Eq.(5), it requires $\mathrm{Im}\tilde{\omega}_{d_y/m_x} = (\tilde{\alpha}_{d_y/m_x})^2 + 2(\tilde{\alpha}'_{d_y/m_x})^2$. A quasi-solution may be found from Eq.(10) when $2(\tilde{\alpha}'_{d_y/m_x})^2$ and $(\tilde{\alpha}_{d_y/m_x})^2$ are comparable. This establishes a relation between electric fields radiated into side ports, as compared with normal ports. Further discussion of its influence on large angle diffraction is available in the Supporting Information. There, we show that an electric or magnetic mode of a higher order (i.e. more spatially dispersive in the lateral dimension) will be desired when it demands a higher field in the +1$^{st}$ diffractive orders for efficient diffraction at large angles. It shifts dramatically the frequency of the corresponding mode, which would challenge the aforementioned phase relation between EM resonances.

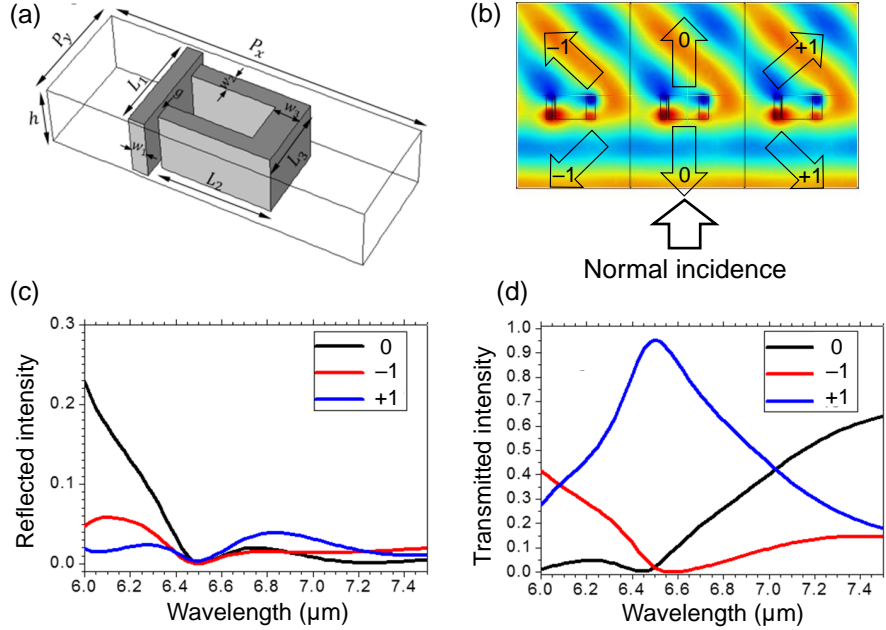

Figure 2. (a) A schematic view of a unit cell consisting of the Si '| ]'-shaped antenna in a vacuum. Here, the geometric dimensions are $P_x = 9.20$ μm, $P_y = 3.20$ μm, $L_1 = 3.00$ μm, $L_2 = 3.25$ μm, $L_3 = 2.18$ μm, $w_1 = 0.47$ μm, $w_2 = 0.48$ μm, $w_3 = 0.76$ μm, $g = 0.17$ μm, and $h = 1.92$ μm. (b) In the simulation, an incident beam impinges normally on the metasurface. Three diffraction orders are present at each side of the metasurface. The color plot of $E_y$ indicates a near-perfect diffraction into the +1st order. (c) Spectra of efficiency for backward scattering diffraction orders. d) Spectra of efficiency for forward scattering diffraction orders.

**Numerical design and optimization of the metagrating.** CMT parameters depend on the metamolecule geometry. For example, the electric mode $d_y$ and out-of-plane magnetic mode $m_z$ are sensitive to the transverse dimensions, while $m_x$ and $m_Q$ modes also strongly depend on the thickness of the metagrating. In addition, the overlap of modal profiles with +1st diffractive plane waves is modified by the transverse spread of metamolecules. While $d_y$ and $m_x$ modes can be accessed under normal incidence in non-bianisotropic structures, it is the symmetry breaking about $yz$-plane that allows access to otherwise dark $m_Q$ and $m_z$ modes. As far as specific design is concerned, we use a previously utilized silicon '| ]'-shaped metamolecules that were, for instance, shown to provide near-$2\pi$ phase coverage and unity transmittance [24]. This geometry is a variation of the 'Γ|'-shaped metamolecule that has been studied on a plasmonic system for Fano resonances due to a coupling between

$d_y$ and $m_z$ [46] [49] [50] and an all-silicon metasurface where higher order electromagnetic resonances, including the $m_x$ and $m_Q$, were probed [4].

Our current design can begin with two parallel *y*-antennae or, alternatively, with a square ring structure (Fig.1b). Two *x*-antennae extrude symmetrically at both ends of one of the *y*-antennae, in order to prevent polarization conversion. The selection of geometric parameters is based on our previous successful tuning of complex mode frequencies of a pair of electric and magnetic modes to achieve a $2\pi$ phase coverage for unity transmissions. Here, we need to operate additional modes to access all orthogonal diffractive components. Given the energy conservation condition and an arbitrary phase assumption for the unity diffraction, the 6 complex linear equations in Eqs.(6) requires at least 10 mutually independent modal parameters. In our parametric search, however, only 9 different dimensions were allowed to be variables during the optimization process, which are illustrated in Fig.2(b), except $P_x$. This is sufficient for achieving an optimized diffraction efficiency of greater than 95%, using the silicon '| |'-shaped design as shown in Fig.2. Given that the operation wavelength is chosen at 6.5 μm and the period of the unit cell in the *x* direction is $P_x$ =9.2 μm, the first diffraction order radiates at an angle of 45°. In principle, the operating frequency, diffraction angle, and the period in *x*-direction are application-dependent and can be flexibly alternated. The resulting diffraction spectra for an optimized structure are presented in Fig.2(c) and (d). Under a normally incident beam, diffraction into the +1st order peaks at the designed wavelength with a 95% efficiency, whereas all other channels are effectively suppressed. In Supporting Information, we present a blazed grating that diffracts at 70° with >93% efficiency for $P_x/\lambda = 1.065$. For efficient diffraction of an *x*-polarized light, it is intuitive that if our current CMT model is rewritten using magnetic field components and electric modes exchange their status with magnetic modes, the theoretical limit of an efficiency can still be near perfect. For this case, we also numerically explored the corresponding symmetry breaking that is available on a dielectric metamolecule and diffraction efficiency in the Supporting Information.

**Experimental verification.** We verify the blazed metagrating concept experimentally by fabricating and testing nanostructures based on a silicon-on-insulator wafer. An undoped, double-side-polished wafer (Ultrasil, Inc.) with a device layer of 2.7 µm and a buried oxide layer of 1.0 µm was cleaned in acetone and isopropyl alcohol (IPA) and coated with two layers of positive electron beam resist: 100 nm of PMMA 950 on top of the 200 nm of PMMA 495k, baking the resist at 170 ºC for 15 min after each coating step. The desired pattern was exposed over the substrate at 1000 µC/cm$^2$ (JEOL 9500FS) and developed in MIBK:IPA 1:3 solution for 75 sec, with consequent rinsing in IPA. A 60-nm-thick Cr hard mask was deposited using electron beam evaporator at a rate of 3 nm/min. After liftoff in sonicated acetone (60 sec), HBr plasma dry etching of silicon down to the 1-µm-thick buried oxide layer was carried out (Oxford Cobra), leaving the desired pattern carved in the device layer. As the last step, the residual Cr mask was removed with a Cr wet etch. The patterned area of each metagrating was (1.5 mm)$^2$. A scanning electron microscope image of the best-performance sample is given in Fig. 3(a). The sample has the same symmetries with that shown in Fig. 2(b); however, a different set of dimensions was found to be the optimum one, owing to the Si–SiO$_2$ substrate present in the experiment.

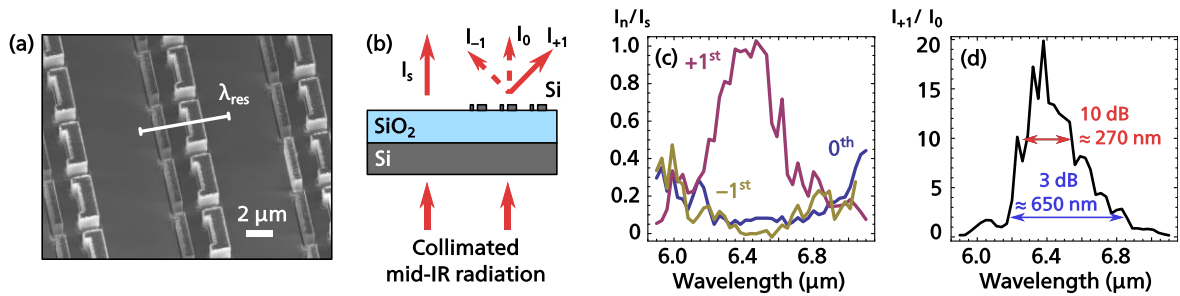

Figure 3. (a) Scanning electron micrograph of the best-performance metagrating array etched in the device layer of a silicon on insulator wafer. $\lambda_{res}$ measures the wavelength of operation, 6.4 µm. (b) Experimentally, the metagraing deflects all the radiation that would have been transmitted by the substrate, to the +1$^{st}$ diffraction order. $I_0$, $I_{+1}$, $I_{-1}$, and $I_s$, are the intensities of the 0$^{th}$, +1$^{st}$ and –1$^{st}$ transmitted diffracted orders, and the signal transmitted through the bare substrate, respectively. (c) Diffraction efficiency spectra of the +1$^{st}$ (purple), 0$^{th}$ (blue) and –1$^{st}$ (yellow) transmitted diffraction orders. (d) The ratio between blue and purple curves ($I_{+1}/I_0$) in panel (c) showing efficient blazed grating operation. The lower boundary for $I_{+1}/I_{-1}$ at 6.4 µm is around 50, limited by the noise of the detection system.

The diffraction efficiency of the fabricated metagratings was quantified using a setup based on a quantum cascade laser (Daylight Solutions MIRcat) as a source of highly collimated monochromatic tunable quasi-CW mid-IR radiation, as depicted in Fig. 3(b). In the experiment, radiation was softly focused with a 200-mm-focal-length lens to a spot 800 µm in diameter, from the back side of the wafer. The intensity of diffracted radiation was measured using a pyroelectric array camera (Ophir Spiricon Pyrocam III) by integrating the intensity over the whole array. For each wavelength, the intensities of the $+1^{st}$, $-1^{st}$ and $0^{th}$ transmitted diffraction orders ($I_{+1}$, $I_{-1}$, and $I_0$) were detected, and the $1^{st}$ and $-1^{st}$ orders were also detected in reflection. The transmission diffraction efficiencies of the best-performance sample are shown in Fig. 3(c). The reflected diffraction orders were beyond detection limit and represented a minute correction over the efficiency of the +1st order, which demonstrates near-perfect efficiency near the design wavelength of 6.4 µm. Note that, since a substantial (~50%) amount of light was experiencing reflection from the high-contrast Si–air and Si–SiO$_2$ interfaces, we were normalizing the intensity of the diffracted beams by the intensity of the beam passing through the unstructured part of the wafer, where the device layer is etched away. In Fig. 3(d), the ratio between the diffracted ($I_{+1}$) and transmitted ($I_0$) intensities is given, reaching 15–20 near the design wavelength. The ratio between $I_{+1}$ and $I_{-1}$ cannot be reliably defined near the design wavelength, as $I_{-1}$ is below the detection limit of our setup; the lower boundary for $I_{+1}/I_{-1}$ at 6.4 µm is around 50, which ensures efficient unidirectional diffraction. Note that efficient diffraction occurs for a fairly broad set of mid-IR wavelengths, showing 270 nm of 10 dB rejection and 650 nm of 3 dB rejection.

**Truncated metasurfaces.** Using the single-metamolecule design approach, our metasurface can operate even after being truncated to small patches. The same design assembled in 5x11 and 3x7 metamolecule arrays was numerically tested under illumination by a tightly focused Gaussian beam. Fig.4(a) shows the setup of the simulation and a wave propagation for the blazing wavelength. At $\lambda = 6.56$ µm, decomposition of the transmitted far fields within the elliptical boundary in Fig.4(b) on the interface reveals that the energy flow is confined within a small range of angles with a central angle of about 45º. The efficiency of this diffraction beam is significantly higher than all the others, while the figure of merit (FOM), which is defined as the diffraction efficiency at a single port normalized to the total energy transmitted in the forward direction, is found near 75% with a slightly red-shifted peak after the energy flow is integrated within a 20º by 20º region around the designed blazing angle. For the sake of comparison, we numerically designed an 8-step phase gradient blazed grating and tested its truncated version. As shown in the Supporting Information, the latter provides the similar diffraction efficiency, being, on the other hand, thicker (4.1 μm) and more demanding in terms of nanofabrication (high aspect ratio). This result suggests single resonator design strategy is effective and can be further optimized for elements working in a small group. On the application side, our metagratings can serve as a basis for novel flat optics components that require small footprint and operate in compact spectrometers, beam deflectors, polarimeters and other devices.

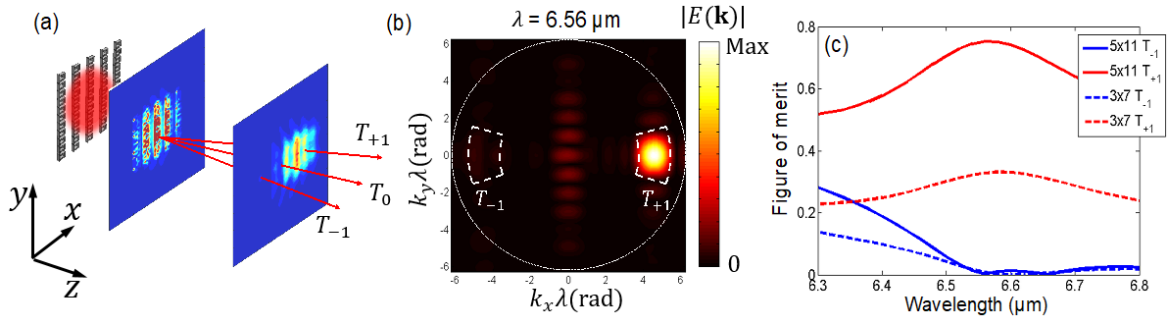

Figure 4. (a) Diffraction of tightly focused Gaussian beams by a truncated bianisotropic metasurface. Color maps show characteristic near-field patterns of the diffracted beams. Arrows indicate diffraction orders. Half width at $1/e^2$ of the Gaussian beam is $3\lambda$. (b) Far-field decomposition of the transmitted field at 6.56 μm. Dashed areas indicate the area of integration of power flux. (c) Transmission efficiencies of three diffraction orders represented by power flux integrated over the dashed areas in (b). The solid red (blue) line plots the FOM of the +1$^{st}$ (−1$^{st}$) diffraction order for a 5x11 unit cell array. The same quantities are plotted with dashed lines for a 3x7 unit cell array.

In conclusion, we have demonstrated the control of diffraction efficiencies by fundamental electromagnetic modes of a metagrating consisting of bianisotropic metamolecules. Coupled-mode theory calculations show that a minimum of 4 fundamental modes of various symmetries are necessary for a perfect 6-port blazed grating at a fixed frequency and blaze angle. We numerically demonstrate blazed gratings with a near-perfect diffraction efficiency at angles of 45° and so on, enabled by sufficient number of degrees of freedom to access all 6 diffractive components. The bianisotropic approach is verified experimentally by a blazed grating based on a silicon-on-insulator platform for mid-infrared wavelengths, where the high diffraction efficiency is confirmed. This analysis provides a simple strategy for general grating designs based on electromagnetic modes, which can be easily extended to a diffraction grating with any number of ports or potentially to other beam steering devices, to find use in spectroscopy, polarimetry, lidars and other photonics applications.